\documentstyle[11pt,aasms4]{article}

\accepted{Astrophysical Journal, October 1997}
\journalid{}{}
\articleid{}{}

\begin{document}
\title{Radio-Luminous Southern Seyfert Galaxies.
I. Radio Images and Selected Optical/Near-Infrared Spectroscopy}

\author{M. A. Bransford\altaffilmark{1,2}, P. N. Appleton\altaffilmark{1,2}}
\authoremail{mabransf@iastate.edu, pnapplet@iastate.edu}
\affil{Erwin W. Fick Observatory and Department of Physics and
Astronomy, Iowa State University, Ames, IA 50011}

\author{C. A. Heisler\altaffilmark{1}}
\affil{Mount Stromlo and Siding Spring Observatory, ANU, Private Bag,
Weston Creek Post Office,  ACT 2611, Australia}

\author{R. P. Norris\altaffilmark{1}}
\affil{CSIRO Australia Telescope National Facility, P. O. Box 76, Epping, NSW,
2121, Australia}

\author{A.P. Marston\altaffilmark{3}}
\affil{Dept. of Physics and Astronomy, Drake University, Des Moines,
IA 50311}

\altaffiltext{1}{Visiting astronomer at the Anglo-Australian
Observatory} 
\altaffiltext{2}{Visiting astronomer at the Australia Telescope National
Facility. The Australia telescope is funded by the Commonwealth of
Australia for operation as a National Facility managed by CSIRO.} 
\altaffiltext{3}{Visiting astronomer at CTIO, CTIO is operated by
Associated Universities for Research in Astronomy (AURA) under
contract to the NSF}

Send off-print requests to M. A. Bransford e-mail: mabransf@iastate.edu.
Higher quality figures (.ps) are available via anonymous ftp, ftp
pv7408.vincent.iastate.edu, login: anonymous.mabransf, password:
yourusername@domain, cd outgoing, binary, get figures.tar (approx. 1
Meg). Updated version includes revised coordinates of galaxies in Table
1.

\begin{abstract}
This is the first of two papers in which a study is made of a sample of
12 southern radio--luminous Seyfert galaxies. Our aim is to investigate
possible correlations between radio morphology and nuclear/circumnuclear
emission-line properties. In this paper we present radio images at
${\lambda}$ = 13, 6, and 3 cm taken with the Australia Telescope Compact
Array (ATCA), global far-infrared (FIR) properties for the whole sample,
and optical and near-infrared (NIR) spectroscopy of an interesting
subset. We find a mixture of radio morphologies, including linear,
diffuse and compact sources. When the FIR colors of the galaxies are
considered there is an indication that the compact radio sources have
warmer FIR colors than the diffuse sources, whereas the linear sources
span a wide range of FIR colors. There is a wide variation in radio
spectral--indices, suggesting that free--free absorption is significant
in some systems, particularly IRAS 11249-2859, NGC 4507, and NGC 7213.

Detailed emission-line studies are presented of 4 galaxies IC 3639, NGC
5135, NGC 3393 \& IRAS 11249-2859. In IC 3639 we present evidence of
vigorous, compact star formation enclosed by very extended
[OI]~$\lambda$6300 emission, suggestive of the boundary between a diffuse
outflow and the surrounding ISM. In another galaxy, IC 5063, we see
evidence for the possible interaction of a highly collimated outflow and
the surrounding rotating inner disk. Of the 5 galaxies which show
compact radio emission, 4 have radio/FIR flux ratios consistent with an
energetically dominant AGN, whereas IC 4995 exhibits evidence for a very
compact starburst. \end{abstract}

\section{Introduction}

Indications that Seyfert galaxies may contain a significant starburst
component come from several different directions, although it is not yet
clear how such star formation might affect the properties of an active
galactic nucleus (AGN).  Norris, Allen, \& Roche (1988) and Roy et al.
(1997) showed that most Seyfert galaxies, unlike radio quasars, roughly
follow the radio-far-infrared correlation, which suggests that their
bolometric luminosity is dominated by star formation activity. Extended
minor-axis radio emission from Seyferts (Baum et al. 1993) suggests that
many Seyfert galaxies may contain powerful circumnuclear starbursts. The
inner few hundred parsecs of Mrk 477 have recently been shown to contain
an extremely luminous and compact starburst (Heckman et al. 1997). In
NGC 1068 an extended luminous starburst region is observed 3 kpc from
the central AGN (e.g. Telesco \& Decher 1988).  Genzel et al. (1995)
made high-resolution NIR spectroscopic images of NGC 7469, which show a
circumnuclear ring of star formation 480 pc in radius. A blueshifted
ridge of emission between the ring and the nucleus is interpreted as
infalling gas that may provide fuel for the AGN. The star-formation in
NGC 1068 and NGC 7469 may result from a stellar bar which is both
triggering extended star-formation and funneling material inwards to
the AGN. Previous investigations have shown that half of the total
luminosity of NGC 1068 (Telesco et al. 1984; Lester et al. 1987) and Mrk
477 (Heckman et al. 1997), and nearly two-thirds of that of NGC 7469
(Genzel et al. 1995), arises from starburst activity.

This study is aimed at looking for possible correlations between
arcsec radio structure\footnote{\footnotesize At the distance of our
sample galaxies this corresponds to linear scales of a few hundred
parsecs.} and optical--IR emission-line properties in a sample of
southern radio-luminous Seyfert galaxies (see Table 1).  The mapping of
arcsec radio structure in southern Seyfert galaxies has only become
feasible recently with the advent of the Australia Telescope National
Facility and many of these radio sources have not yet been mapped.  Our pilot
study focuses on one aspect of the properties of AGNs, namely the
possible relationship between AGN or starburst activity and the
morphology of the arcsec-scale radio emission.  For example, we wish
to investigate the possibility that radio-emitting plasma ejected from
an AGN might interact and trigger star formation in the inner
regions of galaxies. Alternatively, in other systems, the radio
emission itself may originate in circumnuclear star formation and this
will be investigated via spectroscopic observation. 

In this paper we present the radio images, at three wavelengths, of the
galaxies in our sample, as well as optical and NIR spectra for an
interesting subset. In addition we will compare the thermal properties
of the dust emission (as measured by global FIR properties of the
galaxies, e.g. Heisler \& Vader 1994; Condon et al. 1991) with the
radio properties. We show that for a subset of these data, the FIR color
temperatures seem to correlate with radio morphology, suggesting that in
some cases the FIR colors are intimately connected with the nuclear dust
properties of Seyferts.  (see $\S$4.3). In a second paper (Paper II) we
will complete the spectroscopic observations and draw more general
conclusions about the sample.

In $\S$2 we present the sample selection, in $\S$3 the details of the
observations, in $\S$4 the radio data, and in $\S$5 we present optical
and NIR spectra of some of the galaxies. In $\S$6 we present our summary
and conclusions.

\section{The Sample}

The sample galaxies listed in Table 1 are a radio-bright subset of a 2.4
GHz Australia Telescope National Facility (ATNF) survey of Seyfert
galaxies by Roy et al. (1997).\footnote{\footnotesize The Roy et
al. sample is drawn from the de Grijp et al. 1992 sample of IRAS
selected galaxies.} Galaxies from this list were selected if they were
more luminous than L$_{\rm 2.4 GHz}$ $>$ 10$^{21.7}$ W Hz$^{-1}$, with
declinations $<$ -20$\arcdeg$, and redshifts z $<$ 0.04. One galaxy that
met these criteria, IC 4329A, was not included in our sample since it
had been imaged in the radio by A.L. Roy (private communication) and
radio maps of this source will be published elsewhere. Our 2.4 GHz
observations reveal that the range in luminosity of our sample is
log(L$_{\rm 2.4 GHz}$/W-Hz$^{-1}$) = 21.7-23.2 (see Figure 1).  By way
of comparison, Centaurus A, the most radio-luminous Seyfert galaxy in
the southern hemisphere, has L$_{\rm 2.7 GHz}$ = 10$^{23.8}$ W Hz$^{-1}$
(Bolton \& Shimmins 1973). Approximately 40\% of the galaxies lie on the
''radio-bright'' (see Fig. 1) side of the well-known FIR/radio
correlation (Helou, Soifer, \& Rowan-Robinson 1985) and were therefore
expected to show composite AGN/star formation characteristics suitable
for a study of this kind. The galaxies in the sample have been observed
as part of various optical emission-line surveys (Phillips, Charles, \&
Baldwin 1983; van den Broek et al. 1991; de Grijp et al. 1992). In Table
1 we list some of the optical properties of the sample galaxies.

\section{The Observations} 

\subsection{Radio Observations}

Radio observations were made with the 6 km Compact Array (ATCA) on
1994 July 24 and 25, at  a wavelength $\lambda$13 cm and on 1995 November
19-22, at $\lambda$6 and $\lambda$3 cm. Each source was observed
for typically six 20-min periods spread over 12 h, in order to provide
good coverage of the u-v (Fourier-transform) plane. The total
on-source integration times were between 1.29 and 5.66 h. Unresolved
phase calibrators were observed before and after every target
observation.  Data were taken in two independent linear polarizations
at frequencies centered on 2.378 and 2.496 GHz with a bandwidth 128
MHz for the 13 cm observations. The 6 and 3 cm observations were made
simultaneously, with frequencies centered on 4.798 and 8.638 GHz with
a bandwidth of 128 MHz. Flux densities were calibrated assuming a flux
density for the standard ATCA calibrator PKS B1934-638 of 13, 5.8, and
2.8 Jy at 13, 6, and 3 cm, respectively.

AIPS was used to perform the standard calibration and data reduction. Images
were made and CLEANed using the AIPS routine MX, and then corrected for
primary beam effects. The rms flux densities in source-free areas of the
CLEANed images ranged from 50 to 720 $\mu$Jy beam$^{-1}$. Table 2 lists
the clean beam dimensions and rms noise levels, for each source observed
at all three wavelengths. In the case of brighter sources ($\geq$100 mJy)
we used ``self-calibration'' to provide additional phase corrections.
The resulting dynamic range achieved in most of the images was typically
50-100 at 13 cm and approximately 20-30 at 3 cm, although in most cases the 
images were limited by thermal noise ratio rather than dynamic range. 

Total fluxes for the sources were calculated by blanking all data
below twice the rms noise in the image in a region close to the
source.  Because these observations are optimized for studying the
nuclear region, they are relatively insensitive to extended
emission. Thus our images are primarily of the nuclear region and will
therefore not show any smooth extended structure on scales of tens of
arcsec, and consequently the total radio fluxes quoted may be an
underestimate. Table 3 lists the fluxes of the sources at each of the
three wavelengths and other radio properties. The images have
resolutions in the range 1-8$\arcsec$. The galaxy NGC 5135 was
observed solely at 13 cm, because published 6 and 3 cm VLA images are
available for this object (Ulvestad \& Wilson 1989). Spectral indices
were calculated, in the core regions, for 13 and 6 cm, 13 and 3 cm,
and 3 and 6 cm ($\alpha$(6,13), $\alpha$(3,13), and $\alpha$(3,6) are
listed in Table 2), by restoring the higher frequency images with the
same beam and cell-size as used at the lower frequencies. To calculate
the spectral index accurately, we measured the flux density over
similar regions for each source, at both frequencies.

\subsection{Infrared and Optical Observations}

NIR spectroscopy and imaging were performed at the 3.9 m AAT, in
1994 July and October. The July observations resulted in only 2 h of
useful non-photometric K-band imaging data because of bad weather.
Direct imaging was done using the HgCdTe 128 x 128 array detector of
the Infrared Imaging Spectrograph (IRIS) and a field of view of
1$\farcm$3 x 1$\farcm$3. Although we obtained K-band images for IRAS
18325-5926, IC 4995, IC 5063 and NGC 7213, only NGC 7130 shows
interesting structure (see $\S$5.2).

Attempts to obtain long-slit, NIR spectra during this run were
unsuccessful because of bad weather. During part of the AAT Director's
discretionary time on 1994 October 21-23, non-photometric conditions
allowed us to obtain a series of long-slit, NIR spectra for four
galaxies: IRAS 18325-5926, IC 4995, IC 5063 and NGC 7213.  At f/15 the
image scale was 0$\farcs$61 pixel$^{-1}$ and the spectral resolution was
$\sim$300. The telescope was programmed to step the slit across the
galaxy in 0$\farcs$6 increments and integration times of 180-220 s were
obtained at each position. In this way, a spectral cube was built up
giving full two-dimensional spectral information across the inner
60$\arcsec$ of each galaxy. Two sets of scans were made, one in each of
the H ($\lambda$1.65 $\mu$m) and K ($\lambda$2.2 $\mu$m) NIR bands.
Infrared emission was detected from the Br$\gamma$, H$_2$ v=1-0 S(1)
2.121 $\mu$m and [FeII]~${\lambda}$1.644 $\mu$m lines for all four
galaxies observed. Representative K-band spectra are presented in $\S$5.

We obtained long-slit optical spectroscopy at both the AAT and CTIO of
four galaxies, to search for extended emission associated with their
radio and NIR structure in their nuclear regions and inner
disks. The spectroscopic optical observations were made at intermediate
dispersion at the AAT in 1995 August and low dispersion at the 1.5 m
CTIO telescope in 1995 April.

The AAT observations were made using the RGO spectrograph, in
combination with the TEK 1k CCD detector (1024 x 1024) at f/8 with the
25 cm camera and a grating with 270 grooves mm$^{-1}$ in first order to
produce a spectral resolution of $\sim$3.3 ${\rm \AA}$ pixel$^{-1}$ over
a wavelength range of 4170-7625 ${\rm \AA}$. The spatial plate scale was
0$\farcs$77 pixel$^{-1}$. The slit was set to a width of 1$\farcs$66 for
galaxy observations, which was comparable to the atmospheric
seeing. Total integration times at each position were 1200-1800 s. These
observations were made during photometric conditions for only 3 hours on
the first night. The standard star LTT 7379 (Stone \& Baldwin 1983) was
observed at approximately the same airmass as the galaxies, with a wide
slit to collect all the light, for purposes of flux calibration. The
rest of the night, and the next night were wiped out because of poor
weather. Two fully flux-calibrated data sets were obtained for IC 3639
and NGC 5135 for various slit orientations and positions.

CTIO spectra were taken for two galaxies (NGC 3393 and IRAS 11249-2859)
in 1995 April with the 1.5 m telescope, in combination with the GEC 10
CCD, at f/7.5. The grating used provided a very low spectral resolution
of 8.5 ${\rm \AA}$ pixel$^{-1}$ and a wavelength coverage of 4325-6750
${\rm \AA}$.  The plate scale was 1$\farcs$9 pixel$^{-1}$. The slit
width was 3$\arcsec$, which was larger than the atmospheric seeing
(typically 1$\farcs$2). The integrations were 900 s in length. The
spectra were obtained before we knew what the radio structure of our
sample was, and so all the slit positions were taken along position
angle (PA) 90$\arcdeg$ (N through E). For both the CTIO and AAT 
spectra we made no correction for
telluric absorption.

The fluxes of the emission-lines were corrected for internal reddening
using the observed Balmer decrement and assuming a standard interstellar
reddening curve (Savage \& Mathis 1979). The relative line emissivities
for Case B recombination were given by Osterbrock (1989). We used the
SPLOT task in IRAF to measure and de-blend the optical features in our
spectra. We extracted spatial information of the emission-line gas from
the various slit positions, by binning the flux into $\sim$1$\farcs$4
increments for the AAT spectra. For the photometric data, the line flux
ratios are accurate to the $\pm$10-15\% level in the brighter nuclear
regions, decreasing to $\pm$20-30\% far from the nucleus. Table 4
presents a complete summary of the observations made on each galaxy at
radio, NIR, and optical wavelengths.
 
\section{The Radio and FIR Properties}

In this section we present our radio observations. In $\S$4.1, we
provide quantitative and qualitative definitions of the L-, D- and C-class
radio morphologies, along with the presentation of the
full-resolution $\lambda$13, 6, and 3 cm contour maps, which are also shown in
Figure 2a,b. In $\S$4.2 we present and discuss the radio spectral
indices derived from our radio data, and in $\S$4.3, we discuss the
FIR/radio correlation and FIR spectral indices, and their role in
distinguishing AGN from starburst activity.

\subsection{Radio Morphology}

The flux densities, morphology, and radio spectral indices derived from
the radio images are presented in Table 3. 50\% of the galaxies observed
at 13 cm show extended emission, and $\sim$ 33\% show extended emission
at both 6 and 3 cm. We find a wide variation in radio spectral
indices. Free-free absorption is likely to be important in the cores of
two systems: NGC~4507 and NGC~7213, and in the outer radio components of
IR~11249-2859 (see $\S$4.2). There is also a wide variation in the radio
morphologies of the galaxies, which fall into four main classes
(e.g. Wilson 1988).  The physical significance of these classes is
supported by the fact that different optical emission line ratios are
associated with the different morphological radio types and that there
appears to be some correlation with IRAS color ratios (see $\S$4.3):

1. {\bf D--class} diffuse radio morphology which is characterized by a
low surface brightness component surrounding a compact core. The low
surface brightness emission is generally extended on scales $\geq$
5$\arcsec$. D-class sources have been linked with starburst activity
in that the brightness temperature and spectral index of the radio
emission are consistent with a combination of supernova remnants,
diffuse synchrotron emission from supernova-accelerated cosmic-ray
electrons, and H II regions (e.g. Ulvestad 1982; Condon et al. 1982).

2. {\bf L--class} linear structure which straddles a compact core
believed to be powered primarily by an outflow from an AGN. The outflow
may, however, also arise from a circumnuclear starburst (e.g. Baum et
al. 1993). We define the radio emission as ''linear'' if the axial ratio
of the linear structure is $\geq$ 3. L-class radio sources often have
optical emission line ratios consistent with high ionization,
high-velocity gas associated with the narrow-line region (NLR) (Haniff,
Wilson, \& Ward 1988; Wilson, Ward, \& Haniff 1988; Whittle et al.
1988), and as such are intimately connected with an AGN. Whittle et
al. (1988) studied the velocity structure of the [OIII]$\lambda$5007
emission line in a sample of L-class sources, and found that in some
cases there was evidence for an outflow of ionized gas, suggesting that
the radio emission of some L-class sources may be due to an outflow of
plasma from an AGN.

3. {\bf C--class} compact sources, which are either unresolved or only
marginally resolved on the arcsec scale by our present observations
(see Table 2). 

The D and L classifications have to be used with care, as they refer to
the dominant source of radio flux which can cover a wide range of scales
(arc seconds to arc minutes). For example, a faint L-class source may be
contained in D-class emission, as in NGC 1068, where a collimated
structure is seen inside a starburst disk (e.g. Wilson \& Ulvestad
1982b; Wynn-Williams, Becklin, \& Scoville 1985). Other ambiguities may
also arise.  A D-class source seen edge-on might appear as an L-class
source. A linear radio structure could be the result of an outflow from
an AGN or circumnuclear starburst, but could also be emission from a bar
or an edge-on spiral arm. These ambiguities can sometimes be avoided by
comparing the optical and radio emission.  For example, radio emission
from a spiral arm or a bar will be coincident with the optical
structure.

As shown in Table 3, our sample contains two D-class, four
L-class, one is a hybrid L+D-class (see below), and five C-class sources.
IC 3639 and NGC 5135, the only D-class sources in our sample, are
nearly face-on galaxies and their D classification was
straightforward and not hampered by any orientation effects.

On the other hand, care must be taken in the interpretation of the
L-class objects in the sample.  In one case, NGC 7130, the elongated
radio structure may be due to the existence of an asymmetric stellar
bar in the galaxy. This source is known to contain vigorous circumnuclear
star-formation (Shields \& Filippenko 1990 - see $\S$5.2), and the
long-dimension of its $\lambda$13 cm radio structure is oriented
along a position angle (PA) 0$\arcdeg$ which is close to that of an
inner bar. We therefore attribute the radio emission in this galaxy to
either shocks or star formation within the bar. 

However, radio emission from the other L-class galaxies probably
represents an outflow from the nucleus. The PA of the linear structure
seen in NGC 3393 at 13 cm is 23$\arcdeg$ and is not coincident with
spiral arms or a bar. The asymmetric double radio source seen at 3 cm
is also seen in recent VLA 6 cm observations at higher spatial
resolution (R. Morganti, private communication).
NGC 4507, while not highly collimated, meets our criterion for an
L-class, although the knots seen at 6 cm to the north and the south
could be due to a ''hot spot'' ring, extra-nuclear H II regions, or SN
near the nucleus, and not a nuclear outflow. As we discuss in $\S$4.2,
the radio spectral index of the nucleus of NGC 4507 is flat,
suggesting that there is significant free-free absorption.
The position angle of the linear structure seen at 13 cm and 6 cm in
IRAS 11249-2859 is 125$\arcdeg$ and is at an angle of 25$\arcdeg$ with
the major axis of the galaxy. It is therefore unlikely that the
emission is from H II regions at the end of an edge on spiral arm or
from a bar.

\subsection{Radio Spectral Indices}

There are four main effects which are likely to influence the radio
spectral index of Seyfert and starburst galaxies:

1.  the synchrotron emission from the interaction of cosmic rays with
the interstellar magnetic field.  The cosmic rays are generated and
re-accelerated by supernovae and supernova remnants, but have a lifetime
which extends their range beyond the supernova remnants themselves. This
synchrotron radiation is expected to have a spectral index of $\sim$
-0.7. The emission generally dominates the radio power of starburst
galaxies, and is presumably responsible for the tight FIR/radio
correlation obeyed by these galaxies;
 
2.  the synchrotron emission from relativistic particles generated by
the AGN. This radiation could, in principle, cover a wide range of
spectral indices. For example, the cores of radio-loud radio galaxies
and quasars typically have a flat-spectrum core (because of synchrotron
self-absorption) but steep-spectrum extended radio-lobes (because of the
cooling of high-energy electrons). However, most Seyfert galaxies have
AGN core spectral indices in the region of -0.7. Condon et al. (1991)
argue that, amongst infrared-luminous galaxies, even those that are
flat-spectrum do not have a sufficiently high brightness temperature for
synchrotron self-absorption to be important;

3.  thermal emission from H II regions. Compact H II regions in our
galaxy generate free-free emission from hot electrons. Most are
optically thin, giving a flat spectrum, but most compact H II regions
are optically thick at centimeter wavelengths, giving a spectral index
$\sim$2. However, Condon et al. have demonstrated that, at centimeter
wavelengths, the integrated flux of such regions is generally small
compared to the synchrotron emission, so that star-formation regions in
starburst and Seyfert galaxies have a typical synchrotron spectrum;

4.  the radio emission from ultra-luminous infrared galaxies is optically
thick to free-free absorption, so that the typical synchrotron spectrum
of these galaxies is flattened at low frequencies (Condon et al. 1991).

The combined result of these effects in Seyfert and starburst galaxies
is to produce a typical radio spectral index of -0.7 with a flattening
at low frequencies in some starburst sources because of free-free
absorption.  In our sample of galaxies, excluding NGC 4507 and NGC 7213,
the spectral indices of the cores lie in the range -0.57 to -1.05, which
is consistent with a dominant non-thermal synchrotron process, although
the cosmic-ray electrons may be associated either with star formation,
or else with an AGN core.

NGC 4507 and NGC 7213 differ from the other Seyferts in our sample,
having low spectral indices ($\alpha$(3,6) = -0.34 and -0.25,
respectively).  Other Seyfert galaxies are known to contain
flat-spectrum radio sources (Ulvestad \& Wilson 1989). These authors
suggested that the flat spectral indices they derived for some of their
galaxies could arise from synchrotron self-absorption or free-free
absorption from the ionized gas. Given that NGC 7213 is a Seyfert 1
galaxy, it is relevant that Antonucci \& Barvainis (1988) reported that 8
out of 17 RQQs and Seyfert 1 galaxies imaged at 20, 6 and 2 cm had flat
spectral indices, arising from a flat or rising component which becomes
dominant by 2 cm. This 2 cm ''excess'' could be due to free-free
emission either from a starburst (requiring a brightness temperature
$\leq$10$^{\rm 4}$ K) or from $\sim$10$^{\rm 6}$ K nuclear free-free
emission (Antonucci \& Barvainis 1988).

The core of NGC 7213 is marginally resolved by our 1$\arcsec$ beam at 3
cm. Assuming a core size equal to our synthesized beam, the brightness
temperature of the core in this source is about 3000 K, which is too low
for synchrotron self-absorption, suggesting that free-free absorption is
responsible for the low spectral index, although free-free emission is
also plausible. The free-free absorption may arise from a vigorous
starburst present in the nuclear region of this galaxy, or, perhaps,
from a high column density narrow-line region. Interestingly, NGC 7213
contains a ring of H II regions 20-40$\arcsec$ away from the nucleus
(Evans et al. 1996), which, at the distance of this galaxy (see Table
1), corresponds to a ring 2-3 kpc in radius. Yet, our radio observations
reveal that in the inner 1$\arcsec$ (or inner 100 pc) star-formation is
an important contribution to the radio emission. Radio observations of
much higher resolution and dynamic range are important for this source
in order to understand the morphology of the radio emission.

Unfortunately, the core of NGC 4507 is unresolved at 3 cm so we cannot
calculate the brightness temperature of the emission for that source,
and as a consequence we cannot rule out synchrotron self-absorption as
the cause for the flat spectral index.

We have obtained radio spectral indices of the knots in the linear
structure imaged at 13 and 6 cm, in IRAS 11249-2859. The spectral index,
$\alpha$(13,6), of the first knot to the south-east (see Fig. 2a,b) is
-0.88, consistent with synchrotron emission, whereas the first two knots
to the north-west are -0.23 and -0.34, respectively, which is consistent
with free-free absorption.

\subsection{FIR/Radio Correlation and FIR Spectral Indices}

The FIR/radio correlation is thought to arise from a common dependence
of the FIR and radio emission on the formation of massive stars
(e.g. Lisenfeld, Volk, \& Xu 1996), and it is generally believed to hold
up to high FIR and radio luminosities.  Galaxies, or regions within a
galaxy, that adhere to the FIR/radio correlation are likely to be
regions of massive star-formation.  Our sample contains a mix of
galaxies that fell on, or were radio-bright with respect to, the FIR
correlation, as illustrated by Figure 1, which is based on our 13 cm
flux densities (see Table 3) and published IRAS fluxes (see Table 5).

A comparison between the FIR flux and the radio flux is a useful
indicator of the dominant FIR and radio emission mechanisms.  For
example, Helou et al. (1985) define a parameter q, which is defined
as:

\hspace{.5in}q $\equiv$ log[(FIR/3.75 x 10$^{12}$ Hz)/S$_{\rm 1.49
GHz}$]\hspace{.5in}(1)

\hspace{.5in}where FIR $\equiv$ 1.26 x 10$^{-14}$(2.58S$_{\rm
60\mu}$+S$_{\rm 100\mu}$).\hspace{.24in}(2) 

It has been shown that q is an indicator of the relative importance of
starburst to AGN-dominated activity in the nuclear heating of dust in
galaxies. For example, the mean value of q for the IRAS Bright Galaxy
Sample starbursts is 2.34 ($\sigma_{q}$ $\sim$ 0.19) and for AGNs q $<$ 2
(see Condon et al. 1991). 

We present in Table 5 the q-values for our sample, based on IRAS FIR
fluxes and on our 13 cm (2.437 GHz) radio flux densities. We use q
$\equiv$ log[(FIR/3.75 x 10$^{12}$ Hz)/S$_{\rm 2.4 GHz}$],
similar\footnote{\footnotesize Note that we have used a slightly
different definition of q from Helou et al. (1985).  Assuming a spectral
index between 2.4 and 1.49 GHz of -0.7, the borderline between AGN and
Starburst would be shifted by 0.14 in dex between the two definitions,
which is a small effect} to that of Helou et al.~(1985).  Also presented
in Table 5 are the published IRAS FIR fluxes (de Grijp et al. 1992) and
the corresponding FIR spectral indices ($\alpha$ $\equiv$ d log S/d log
$\nu$). Dust heated primarily by a luminous buried AGN is found in some
galaxies, (i.e., the so called ``warmer'' galaxies) and Vader et
al. (1993) suggested that such sources emit strongly near 60 $\mu$m,
whereas the UV emission from a pure starburst population is believed to
be re-radiated mostly between 60 and 100 $\mu$m.  In particular,
galaxies with star-formation tend to have steeper far-infrared spectral
indices ($\alpha$(25,60) $\leq$-1.4 and $\alpha$(60,100) $<$ -1.1), and
those systems dominated by an AGN have a relatively flat FIR spectrum.

Of the 5 ''compact'' (C) radio sources in our sample, 4 (IRAS
11215-2806, IRAS 13059-2407, IRAS 18325-5926, \& NGC 7213) have FIR flux
ratios and q-values that support AGN-dominated FIR and radio
emission. These systems probably have very little nuclear/circumnuclear
star-formation.

The one exception to this is IC 4995. This galaxy shows unresolved
structure in the 3 and 6 cm radio maps and yet has a q-value suggestive
of a star formation--dominated galaxy. Spectra, which will be presented
in Paper II, show little evidence for significant powerful large--scale
star formation in this galaxy which argues against the FIR flux coming
from star formation in the outer disk. Furthermore, the point-like
nature of the K-band image of this galaxy (not shown) also argues
AGAINST disk-wide star formation contributing to the large FIR flux. We
therefore conclude that circumstantial evidence points to IC 4995
containing a very compact starburst region which is unresolved by our
present observations. Alternatively, IC 4995 may turn out to be the
exception to the rule discovered by Helou et al. (that high q values are
indicative of starburst-dominated systems). For this reason it would be
highly advantageous to obtain higher resolution optical, IR and radio
observations of this galaxy.

In Fig.~3, we present a plot of the IRAS S(100)/S(12) flux ratio versus
S(60)/S(25) ratio to create a color-color diagram for the sample.  This
combination of IRAS flux ratios has recently been found to be well
correlated in samples of Seyfert 2 galaxies (e.g. Heisler, Lumsden, \&
Bailey 1997) and is being modeled by Dopita et al. (1997). It appears
that our galaxies also show a tight correlation in these colors. A lower
limit for the S(100)/S(12) flux ratio is shown for NGC 3393, IRAS
13059-2407, and IC 4995. For IRAS 11215-2806, only the S(60) and S(25)
band fluxes are known. Reasonable limits on the S(100)/S(12) flux ratio
would imply that it too follows the correlation (see vertical line in
Fig. 3).

Figure 3 shows a number of interesting properties for the sample. If the
radio morphologies of the galaxies are taken into consideration it is
clear that the C-class galaxies occupy the lower-left (warmer FIR
colors) part of the diagram, whereas the D-class sources (and the one
partially resolved C-class source) lie at the cooler end of the
correlation. The L-class sources span the entire range, although the two
most highly collimated sources, IC 5063 and IRAS 11249-2859 lie in the
warmer region of the diagram. This morphological segregation, based on
IRAS colors, suggests that the global FIR properties are linked in some
way to the radio morphologies.

To understand how this might be, we will consider what has been learned
recently about galaxies with small (i.e. warm) S(60)/S(25) IRAS flux
ratios. Heisler et al. (1997) have shown that Seyfert 2 galaxies with
warmer S(60)/S(25) colors are most likely to show broad permitted lines
in their polarized nuclear light (Hidden Broad-line regions-HBLRs), and
lower line-of-sight extinction to the nucleus.  Within the context of
the unified theory of Seyfert galaxies, Heisler et al.  provide evidence
that the warmer colors are a result of the viewing orientation of the
inner dust torus which is postulated to surround the AGN. A dust torus
seen closer to face-on would show warmer IRAS colors than those seen
more nearly edge-on because the viewer sees more deeply into the inner
regions as a result of a higher viewing inclination.

Can the same argument be applied to our sample? IC 5063 (L+D) shows the
warmest colors in Fig. 3 and is already know to be a HBLR galaxy, and so
its warm colors are consistent with the Heisler et al. picture.  A more
challenging problem is to explain why the C-class sources should have
warmer IRAS colors. One possibility might be that the compact sources
are collimated ``jet-like'' sources seen end-on. If the jets were
emitted roughly orthogonally to the dust torus, then this would provide
a natural explanation for the warmer colors since the viewing direction
would be favorable for seeing deeper into the dust torus. In at least
one of the C-class sources (IR 1832-59) the galaxy exhibits high
excitation optical emission lines (Isasawa et al. 1995) and is
suspected, from X-ray evidence, to contain a Seyfert 1 nucleus. This
would seem to support the view that we can almost view directly the
Seyfert 1 core.
 
Certainly the two galaxies with the coolest colors, NGC 5135 and NGC
7130 show evidence for large-scale extended star formation and their
FIR colors are likely to be dominated by heat sources spread over a
large scale. 

However, we caution that our understanding of correlations like that in
Figure 3 is still in its infancy, and any possible correlation between
warm IRAS colors and radio morphology may be entirely coincidental.
Work such as modeling by Dopita et al. (1997) may help to clarify the
nature of such a correlation in terms of optical depth effects in the
dusty torus at 12 $\mu$m.

\section{Optical and Near-Infrared Data and its Relation to Radio 
Structure}

In $\S$5.1 we present the AAT and CTIO optical spectra, and we discuss
the correlations of the line ratios with radio structure for IC 3639,
NGC 5135, NGC 3393, and IRAS 11249-2859.  In $\S$5.2 we present the AAT
NIR spectra and imaging, and in $\S$5.3 we discuss in detail the radio
structure of IC 5063 and its relation to published optical imaging and
spectroscopic observations.

\subsection{Optical Spectroscopy of a Subset of the Sample}

\subsubsection{IC 3639}

IC 3639 has been observed spectroscopically and has been imaged in the
the H$\alpha$ line (van den Broek et al. 1991; known as galaxy 24SW in
that paper) as part of a study of galaxies with large FIR-to-optical
luminosities. This Seyfert 2 galaxy shows considerable extended
H$\alpha$ emission over much of its compact, high surface-brightness
disk.  The overall extent of the disk is only 3 kpc. A nearby companion
shows a ring of H II regions (van den Broek et al. 1991; galaxy
24NE). IC 3639 has been shown, from optical spectropolarimetry, to
contain a hidden broad line region (Heisler et al. 1997). Our 13 cm
contour map of this galaxy shows a very rich structure. Surrounding the
central core is a diffuse radio structure with several small unresolved
knots of emission, some of which are coincident with H$\alpha$ knots,
particularly in the northern half of the galaxy. This galaxy is clearly
a D-class radio source. The overall envelope of the radio emission down
to the level of 0.16 mJy beam$^{-1}$ follows closely the boundaries of
the optical galaxy, suggesting that the radio emission is intimately
related to disk processes, i.e., star-formation and supernova events.

Our spectroscopic observations from the AAT reveal highly extended
[OI]~$\lambda$6300 and H$\alpha$ emission. We were able to map spatially
the optical emission-lines in and near the Seyfert nucleus.
Representative spectra are presented in Figure 4a,b,c. In Figure 5, we
plot the line ratios obtained for IC 3639 on the standard optical
diagnostic diagrams (Baldwin, Phillips \& Terlevich 1981; Veilleux \&
Osterbrock 1987) [OIII]~$\lambda$5007/H$\beta$ vs
[SII]~$\lambda\lambda$6731,6716/H$\alpha$, [OI]~$\lambda$6300/H$\alpha$
and [NII]~$\lambda$6584/H$\alpha$ for PA 90$\arcdeg$, centered on the
nucleus. The line ratios change from AGN-dominated emission in the
nucleus, to H II-region-like emission at a radius of $\sim$1.5 kpc,
which suggests that circumnuclear star-formation is important. Note the
very blue continuum in Figure 4c, which suggests a very young stellar
population. Tables 6a,b,c contain emission-line fluxes and flux ratios
for the detected lines at 1$\farcs$4 intervals from the center for the
three observed slit positions (see below).

It is interesting that in two of the line diagnostic diagrams, those
involving ratios of [OI]~$\lambda$6300 and [SII] to hydrogen (Fig.
5b,c), the points representing emission just outside the nucleus
encroach significantly into the LINER region. We will argue below
that this may be a result of enhanced [OI] and [SII] emission caused by
weak (150 km s$^{-1}$) shocks in the circumnuclear environment (Dopita
\& Sutherland 1995).

A marked east/west asymmetry exists in both the emission
lines and the radio distribution (e.g. in Fig. 5a,c, note how the line
ratios show this east/west asymmetry). We will discuss this in relation
to a simple model of the galaxy below.

Along PA 0$\arcdeg$ (see Fig. 6a,c) the line ratios change smoothly
from AGN to H II-region-like emission with increasing distance from the
nucleus. Again we note that the [OI] emission-line ratios dip
significantly into the LINER region. Unlike the spectrum taken at PA
90$\arcdeg$, there is no noticeable asymmetry on either side of the
nucleus.

We also obtained a spectrum 7$\farcs$5 N of the nucleus along PA
90$\arcdeg$ (see Fig. 7a,c for BPT and VO diagrams). The [OI]
distribution is remarkably asymmetric, showing strong LINER-type
emission in the east, but H II-region-like emission in the west. This
mimics the behavior at PA 90$\arcdeg$ through the nucleus, except that
no AGN emission is seen.

The radio emission is highly extended and asymmetric at 13 cm, being
stronger and more extensive to the north-west (see Fig. 2a). The two
spectra which probed the east-west character of the galaxy also reflect
an asymmetry. Strong optical emission on the western side of the
nucleus, in the region of the high surface-brightness radio emission,
shows H II-region-like spectra, whereas to the east the emission is more
characteristic of LINER emission, and this occurs on the edge of the
rapidly declining radio emission. This behavior is seen in both
east-west spectra which cut through the AGN and to the north of the
AGN. We note that LINER characteristics (especially enhanced line ratios
of [OI]~$\lambda$6300 and [SII]~$\lambda\lambda$6716,6731 to H$\alpha$)
could arise from either a low-powered AGN illuminating ambient
narrow-line region gas or from high-velocity shocks (See Dopita \&
Sutherland, 1995 for discussion). The fact that the spectra appear H
II-region-like over the area of the extended radio emission region but
LINER-like at the edges of the radio structure, suggest to us that the
LINER emission may be from shocked gas resulting from a tipped outflow
from a widespread starburst in the central few kpc. The morphology of
the radio emission itself is reminiscent of outflows such as those seen
in starburst systems like M82. Unfortunately, because the radio surface
brightness of the very extended 13 cm emission is low, it is not
possible to determine a spectral index of the plasma, but the radio
emission may be from relativistic electrons which have burst out of the
disk asymmetrically, perhaps from supernova explosions. In this simple
picture, gas from the outflow interacts with ambient ISM material to
shock and produce a LINER spectrum. If the LINER spectrum were
interpreted as coming predominantly from shock waves, they would imply
supersonic flow velocities of the order of 150-200 km s$^{-1}$ for IC
3639 (Dopita \& Sutherland 1995). One can test this model further by
obtaining higher dispersion spectra to look for mass-motions in the gas
in order to identify conclusively a superwind in this galaxy.

An indication that nuclear star-formation may have been episodic in IC
3639 comes from the marginal detection of Balmer absorption (H$\beta$)
in a ring at about 7$\arcsec$ from the nucleus. Three 1-d spectra, at
that angular distance from the nucleus, hint at Balmer absorption.
However, it is not seen in any of the other extracted 1-d spectra. If
these features are confirmed (e.g. by deeper spectra in the blue) this
might suggest evidence for a post-starburst population some 1-3 Gyr old.
The H$\beta$ emission line, with faint absorption superimposed, was not
corrected for absorption and we caution that the A$_{\rm v}$ derived
from the Balmer decrement may be unreliable. As a consequence the line
flux ratios, in these three spectra, could be in error by $\sim$
30\%. We show this feature in Figure 3b.

\subsubsection{NGC 5135}

NGC 5135 is a barred spiral. The 13 cm radio image (see Fig. 2a) shows
considerable extended structure. Our observations show that, in addition
to the central extended emission regions, there are knots of emission on
a larger scale.  In particular, there are two relatively bright knots of
emission, one at a position angle of 10$\arcdeg$, 18$\farcs$4 (4.7 kpc)
distant from the central source, and another at 279$\arcdeg$,
13$\farcs$8 (3.5 kpc) away from the center of emission. Comparison with
the optical image shows that the strongest emission is located at the center of the
bright optical bar (see Fig. 2a).  Ulvestad \& Wilson (1989) obtained 6
and 20 cm images of NGC 5135, providing arcsec resolution with the A and
B arrays of the VLA. At both wavelengths they saw an asymmetric
structure consisting of a bright core with amorphous, fainter extension
to the north-east (D-class source). The extension broke up into several
smaller knots giving it a bubble-like appearance. All of this structure
lies within the inner contours of our 13 cm map.

We obtained optical spectra at the AAT for this galaxy along PA 19$\arcdeg$
and 45$\arcdeg$. The former intersects the radio knot to the north-east
and the latter follows the diffuse radio emission detected by Ulvestad
\& Wilson. Along both position angles, the emission is extended over
$\sim$5$\arcsec$. The optical diagnostic diagrams (see Fig. 8a,c for
representative spectra along PA 19$\arcdeg$ and Fig. 9a,c for BPT and VO
diagrams; also see Table 7 for fluxes and flux ratios) along PA
19$\arcdeg$ reveal that the emission is that of an AGN to the
south-west, but changes in the north-east (where one of the radio knots
resides) from AGN to LINER to H II region-like. Although not very
extended, the optical emission is consistent with star-formation in the
circumnuclear region. Again, this confirms that a D-class source
has significant amounts of star-formation.  Unfortunately, the
observing conditions for NGC 5135 became poor, and it is possible that
fainter off-nuclear emission was not detected.

\subsubsection{NGC 3393}

A CTIO spectrum was obtained for NGC 3393 along a position angle of
90$\arcdeg$ through the nucleus.  Because of the very low dispersion, we
were unable to de-blend the H$\alpha$ and
[NII]~$\lambda\lambda$6548,6584 lines, and so were unable to construct
line ratio diagnostics in the nuclear regions. In both the east and west
directions, [OI]~$\lambda$6300 is extended over $\sim$2$\arcsec$ which,
at the distance of this galaxy, is $\sim$500 pc. The H$\alpha$ is more
extended, covering $\sim$10$\arcsec$, or $\sim$2.6 kpc, to the east and
$\sim$6$\arcsec$, or $\sim$1.6 kpc, to the west, with a similar
morphology to that of the 6 and 3 cm radio emission. The extracted 1-d
spectrum for the nucleus of NGC 3393 is presented in Figure 10a.

\subsubsection{IRAS 11249-2859}

A spectrum was obtained for IRAS 11249-2859 along a position angle of
90$\arcdeg$ through the nucleus, cutting through the radio knot to the
east. We show in Figure 10b a 1-d spectrum of the nucleus of IRAS
11249-2859.

It is interesting to note that the H$\alpha$ is extended over
$\sim$6$\arcsec$, or 2.7 kpc, and [OI]~$\lambda$6300 is extended over
$\sim$2$\arcsec$, or $\sim$1.4 kpc. Proceeding east, away from the
nucleus, the [OI] emission at first declines and then increases in
strength $\sim$4$\arcsec$ from the center and then disappears again at
larger distances. This increase in flux at 4$\arcsec$ corresponds to the
position of the south-east radio knot. A similar increase in line
emission at the position of the knot is seen in the (blended) H$\alpha$
\& [NII]~$\lambda\lambda$6548,6584 emission and the
[SII]~$\lambda\lambda$6731,6716 lines. The ratio of [SII] to H$\alpha$
is 0.39, [NII]~$\lambda$6584 to H$\alpha$ is 0.48, and [OI]/H$\alpha$ is
0.04.  Off the nucleus, H$\beta$ and [OIII]~$\lambda$5007 disappear
completely, thus we cannot construct the [OIII]/H$\beta$ ratio which
would better determine if the emission is AGN-like or consistent with a
star-formation region. We note that these ratios are also consistent
with a low to intermediate velocity shock-wave process (Dopita \&
Sutherland 1995). Given that the optical emission lines are coincident
with one of the components of the linear triple radio source (the SE
knot), it is clear that follow-up observations of higher sensitivity
along a PA of 125$\arcdeg$ are highly desirable. We speculate that these
observations might reveal that the nuclear outflow is causing wide scale
shocks in the disk of this galaxy giving rise to the enhanced [OI] and
possibly inducing secondary star-formation, consistent with the flat
radio spectral indices of the knots to the NW.

\subsection{Near-Infrared Spectra and Imaging}

All four galaxies observed spectroscopically in the NIR (IRAS
18325-5926, IC 4995, IC 5063, and NGC 7213; see Table 4) show detectable
emission lines of Br$\gamma$, H$_{2}$, and [FeII]. The nuclear spectra
are presented in Figure 11.

Unlike the other galaxies observed at NIR wavelengths, IRAS 18325-5926
shows no evidence for an underlying stellar continuum (in particular the
CO (2-0) band-head is missing).  This is consistent with significant
obscuration, even at infrared wavelengths, by dust near the nucleus.
Unusually high levels of obscuration in this probable heavily embedded
Seyfert 1 nucleus was also inferred from the optical and X-ray
observations of Isasawa et al. (1995). We note that the lack of the CO
band-head could be due to a stronger continuum from the Seyfert 1
nucleus.

NGC 7130 may be an example of a collisional ring galaxy (e.g. see the
review by Appleton \& Struck-Marcell 1996) formed as a result of a small
companion passing through the disk of a larger system. In the radio, at
13 cm, this galaxy is clearly extended with a linear north-south
structure (L-class source). However, the origin of the linear structure
is unclear since it is coincident with a bar along the same PA.

This galaxy has been studied in detail by Shields \& Filippenko (1990),
who spectrally mapped the nuclear regions of the galaxy. Their work
clearly shows the composite nature of this galaxy, in that the line
intensity ratios vary radially from those indicative of AGNs, close to
the nucleus, to that of H II regions at increasing distances from the
center.

We obtained a K-band continuum infrared image of NGC 7130 (Fig. 12a,b),
but no line data were obtained. In the high-surface-brightness inner
regions, two tightly-wrapped spiral arms emanate from a somewhat
asymmetric bar. The arms almost form a ring or lens component. The
bright knot in the disk of the galaxy is a foreground star. However, the
high-contrast, grey-scale image shown in Figure 12b clearly shows the
existence of a large faint ring of emission which is confirmed by
the Digitized Sky Survey (DSS) plate of the area. A small
companion galaxy lies embedded in this outer ring and Figure 12 also shows
an additional faint plume to the south-west of the nuclear regions which
may be connected to the outer ring.

NGC 7130 has the highest FIR luminosity, log[FIR/L$\odot$]=11.04, of all
the galaxies in our sample. Much of the star-formation that powers
ultra-luminous far-infrared galaxies is believed, in part, to be
triggered by interactions. While this image alone provides no evidence
for the star-formation properties of this galaxy, its morphology alone
strongly suggests that it is interacting.

\subsection{Radio Structure of IC 5063 and its Relation to Published
Optical Spectra and Imaging}

IC 5063 is a Seyfert 2 galaxy which is known to be abnormally radio-loud
(Danziger, Goss, \& Wellington 1981; Roy \& Norris 1997). We measure a
flux density of 747 mJy at 13 cm. This galaxy, along with IRAS
13059-2407, is the furthest from and thus the most ``radio-loud'' with
respect to, the FIR/radio correlation for our sample. At 3 cm the
emission is in the form a linear structure composed of the bright
central nucleus with two distinct knots towards the south-east, along PA
115$\pm$3$\arcdeg$. Morganti, Oosterloo, \& Tsvetanov (1997) have also
imaged this galaxy at 3 cm and our maps agree very well in morphology
and flux. At 6 cm, the remnants of the linear structure can be seen, but
a faint shell of emission is seen extending from the end of the linear
structure, perhaps indicative of a bow-shock. The diffuse emission
observed at 6 cm may result from the interaction of the linear
structure, i.e. ''plasmons'' ejected from the AGN, with the ambient ISM
in the inner disk. This galaxy is a hybrid L+D-class (see Figure 2b),
since it is an L-class source at 3 cm and a D-class source at 6 cm. In
Figure 13, we label the bright central source (at 3 cm) as knot A, the
first knot to the south-east as knot B, and the second, fainter radio
source as knot C. From our K-band spectra (see Fig. 11) we extract a
log[H$_{2}$/Br$\gamma$] ratio of 0.14, which is confirmation of an AGN
in the nucleus (Mouri \& Taniguchi 1992). AGN-dominated FIR emission is
suggested by the FIR flux ratios and the q-value for this galaxy (see
Table 5). 

IC 5063 is fascinating because it may be a merging system which contains
one of the best-defined ionization cones known to date (Colina, Sparks,
\& Macchetto 1991). It is an S0 galaxy with very peculiar dust lanes and
emission-lines which define a cross-shape centered near the AGN. Colina
et al. have suggested that the cross-shape has some of the
characteristics of an ionization cone illuminated by an AGN. To explain
the increase in the excitation of the ionized gas with distance from the
center, they suggested that this may be the result of decreasing density
and a strong metallicity gradient in the central few arcsec.  However,
they could not completely rule out a local source of heating in the
cone. Of special significance is the fact that the orientation of the
linear triple is coincident with the inner edge of the ionization cone
reported by Colina et al. (1991), in just the region where the ionized
gas shows enhanced excitation. Knot B (see Fig. 13), in the 3 cm map,
seems to coincide exactly with the peak in the brightness profile for
both H$\alpha$ and [OI]~$\lambda$6300 emission. The fact that the peak in
the emission line strength was offset from the ``nucleus'' was discussed
at length by Colina et al. but was not explained.  However, these
authors did note that the velocity of the peak differed by 80 km
s$^{-1}$ from the surrounding fainter gas. We will argue below that this
may be a signature of ionized gas being ejected supersonically into the
surrounding ISM in the form of plasmons.

In an H$\alpha$ imaging survey of galaxies with far-infrared spectral
energy distributions peaking near 60 $\mu$m, Heisler \& Vader (1995)
showed that this galaxy has three bright knots along, roughly, but not
exactly, the same position angle as the radio knots seen in our 3 cm
image. It is interesting to note that radio knots A and B have similar
strengths in H$\alpha$. We have recently obtained new IRIS NIR long-slit
spectra along the radio structure seen at 3 cm (to be published
later). The knot identified in the radio image as B contains both a NIR
continuum source and lines of [FeII]~$\lambda$1.644$\mu$m and Br$\gamma$,
with fainter IR emitting knots on either side (this confirms the
existence of the nuclear [FeII]~$\lambda$1.644$\mu$m detected in our
spectra obtained in 1994 October). Thus, the Heisler \& Vader H$\alpha$
imaging and our NIR spectroscopy suggest that the nucleus is knot B,
although this is by no means clear. In particular, the absolute
positions of the optical and infrared data are uncertain by about an
arcsec, and cannot therefore be accurately superimposed on the radio
image.

Figures 13a,b show that there is a slight difference in relative position 
between the H$\alpha$ knots (shown as crosses) and the 3 cm radio knots.
Figure 13a assumes that radio knot B is the nucleus and, hence,
the H$\alpha$ knot and knot B are made to align, whereas figure 13b
assumes that radio knot A is the nucleus. In each case,
the other H$\alpha$ knots are offset from the radio knots.
We therefore speculate,
regardless of whether knot A or knot B is the nucleus, that these
offsets are consistent with plasmons being ejected into a
counter-clockwise rotating disk. The plasmons ram into the ambient ISM
causing enhanced H$\alpha$ emission along their leading edges. We note
that if two ``extra-nuclear'' radio blobs are indeed moving at a minimum
of 80 km s$^{-1}$ relative to the surrounding gas, they would be highly
supersonic relative to their surroundings and might naturally produce
bow-shocks on their leading edge. It is therefore significant that the
6 cm map shows a bow-shock-like morphology ahead of the possible ejected
plasma (such emission might be produced by electrons trapped in tangled
magnetic fields in the bow-shock ahead of the ejected material). Further
radio observations of higher dynamic range and spatial resolution may
help to define this morphology better. Other evidence to support the
view that gas outflow from the AGN is interacting with the ISM of the
inner disk are broad ($\sim$600 km s$^{-1}$), blueshifted absorption
features seen against a disk of neutral hydrogen emission (Morganti et
al. 1997).

We emphasize that we are not suggesting that the overall morphology of
the optical emission-line properties of IC 5063 are driven by the radio
source, but rather that additional heating is occurring along one edge
of the ionization cone that seems to be related to the radio
structure.

\section{Conclusions}

1) Our sample contains a mixture of radio morphologies. There are four
L-class sources, two D-class sources, one L+D-class, and five compact
(unresolved or marginally resolved on the arcsec scale) sources. Four of
the galaxies in the sample (NGC 4507, IC3639, NGC 5135 and NGC 7130)
show evidence for being AGN/starburst composites, and a 5th galaxy (IC
4995) is likely to contain a compact starburst core on a scale smaller
than our arc second (r$<$ 400 pc) resolution based on its IR and radio
characteristics. Further optical properties of the galaxies not studied
optically here will be presented in Paper II.

2.) There is a correspondence between D-class radio morphology and
circumnuclear star-formation. Our optical spectroscopic observations of
IC 3639, luminous in the FIR (log[FIR/L$\odot$]=10.33), show clearly the
importance of circumnuclear star-formation and the composite
AGN/starburst nature of this galaxy. The radio emission has a clear
east-west asymmetry that is seen also in the optical spectra,
particularly in the (highly extended) [OI]~$\lambda$6300 emission
line. The radio knots appear to be contained within a region of highly
extended optical [OI]~$\lambda$6300 emission. Our BPT diagrams reveal
that the emission changes within the inner 1.5 kpc from AGN, through
LINER, to H II-region-like spectra radially away from the nucleus. The
[OI] emission that is detected to the east of the nucleus is consistent
with emission that is being driven by large scale shocks (v $\approx$
150 km s$^{-1}$), perhaps being generated by the outflow of material
from a nuclear starburst. In a similar vein, our optical spectroscopy of
NGC 5135 also clearly demonstrates the importance of circumnuclear star
formation. Along PA 19$\arcdeg$, the emission-line ratios change
radially away from the nucleus, changing from AGN to LINER to H
II-region-like emission.

3.) Some morphological segregation occurs when the galaxies global FIR
colors are investigated. For example, on a plot of IRAS S(100)/S(12)
versus S(60)/S(25), the C-class sources cluster to warmer IRAS colors
than the D-class sources. The L-class sources have a wide range of
color temperatures, but the hottest galaxies are those with well
defined highly collimated radio emission. It is not clear why the
radio morphology and the IRAS colors are related, unless the higher
colors relate to the orientation of the dust torus (see for example
Heisler et al (1997), and Dopita et al. (1997)), and by implication,
the radio jet direction.

4.) Our CTIO spectra for NGC 3393 and IRAS 11249-2859 show that the
radio and optical emission-line gas is extended on roughly the same
scale. In particular we detected, in [OI]~$\lambda$6300, the radio knot
to the south-east in IRAS 11249-2859. NGC 7130, another of the L-class
sources, has been shown in the work of Shields \& Filippenko (1990) to
contain a significant amount of circumnuclear star-formation. The L
classification of this galaxy is probably due to a bar along the same PA as
the radio emission. Because of the lack of optical and NIR spectra we
cannot make any conclusions about the correlation of, or lack thereof,
radio morphology with circumnuclear star-formation for the L-class
sources. Future papers are planned in order to shed light on this;

5.) The radio spectral indices for the core-regions of the sample are
generally in a range consistent with synchrotron emission, either from
the AGN itself or from cosmic rays accelerated by star-formation
activity. Two notable exceptions are NGC 4507 (L-class) and NGC 7213,
whose low spectral indices indicate that they may suffer from free-free
absorption of a thermal source. A likely source for the free-free
absorption in the marginally resolved radio source NGC 7213 might be
circumnuclear star-formation, as suggested by the brightness temperature
of the radio emission. Furthermore, the collimated structure seen in
IRAS 11249-2859 contains knots that are flat-spectrum sources suggesting
that they might be undergoing free-free absorption;

6.) IC 5063, an L+D class radio source, has IRAS fluxes and a q-value
consistent with AGN-dominated emission, although spectroscopic
confirmation of this statement is pending. The radio emission seen at 6
and 3 cm suggest that ``plasmons'' are being ejected from the nucleus
into the disk of the galaxy, giving rise to extensive shocks, possibly a
bow-shock morphology at 6 cm, and significant local heating of one edge
of a well-defined ionization cone seen in optical emission lines. In the
nucleus NIR spectra revealed Br$\gamma$, H$_{2}$ and [FeII] (along with
the CO (2-0) bandhead feature);

7.) Our infrared observations of IRAS 18325-5926, IC 4995, and NGC 7213
reveal H$_{2}$, Br$\gamma$, and [FeII]~$\lambda$1.644 $\mu$m. Also, we
detect the CO ($\nu$=2-0) band-head absorption feature, longward of
2.3 $\mu$m, in IC 4995, IC 5063 and NGC 7213.

Acknowledgments

We are grateful to Russell Cannon for granting us director's time at the
AAT in order to obtain near-infrared spectroscopy presented in this
paper. We would also like to thank the staff at both the Paul Wild
Observatory and Siding Spring Observatory for their very generous advice
and help. M.A. Bransford thanks Sammy for his warm and friendly demeanor
during visits to Australia, and to R. Morganti for very valuable
comments, which helped to improve the paper. We thank STSci/SERC for use
of the digitized sky survey. We would also like to thank the referee for
very valuable comments and suggestions that helped to improve the
paper. The Australia Telescope is funded by the Commonwealth of
Australia for operation as a National Facility managed by CSIRO. This
work is partly funded by an NSF/US-Australia Cooperative Research Grant
INT-9418142. A. Marston was supported by NASA JOVE grant NAG8-264 and a
grant from NASA administered by the American Astronomical Society.

%
%

%
%

\clearpage
\begin{deluxetable}{lcccccccc}
\scriptsize
\tablecaption{The Sample}
\tablewidth{0pc}
\tablehead{
\colhead{Name} & \colhead{RA(J2000)\tablenotemark{a}}   &
\colhead{DEC(J2000)\tablenotemark{a}}   &
\colhead{$m_{\rm b}$\tablenotemark{b}} & 
\colhead{Redshift\tablenotemark{c}}  & \colhead{Distance\tablenotemark{d}} &
\colhead{Scale\tablenotemark{d}} &
\colhead{Type\tablenotemark{e}} & \colhead{Diameter\tablenotemark{b}}
\\
\colhead{} & \colhead{(h m s)} & \colhead{(d m s)} & \colhead{(mag)} &
\colhead{ } & \colhead{(Mpc)} & \colhead{(kpc arcsec$^{-1}$)} & \colhead{ } &
\colhead{(arcmin)}
} 
\startdata
NGC 3393 & 10 48 23.4 & $-$25 09 44.0 & 13.09 & 0.0137 & 54.8 & 0.266 &
Sy 2 & 2.2 x 2.0\nl
IRAS 11215-2806 & 11 24 02.7 & $-$28 23 15.3 & 13.00 & 0.0135 & 54.0 & 0.262 &
Sy 2 & --- x ---\nl
IRAS 11249-2859 & 11 27 23.4 & $-$29 15 27.5 & 14.71 & 0.0234 & 93.6 & 0.454 &
Sy 2 & 1.6 x 0.4\nl
NGC 4507 & 12 35 36.7 & $-$39 54 34.5 & 12.92 & 0.0132 & 52.8 & 0.256 &
Sy 2 & 1.7 x 1.3\nl
IC 3639 & 12 40 52.8 & $-$36 45 22.4 & 13.00 & 0.0125 & 50.0 & 0.242 &
Sy 2 & 1.2 x 1.2\nl
IRAS 13059-2407 & 13 08 42.0 & $-$24 22 58.0 & 16.00 & 0.0141 & 56.4 & 0.273 &
Sy 2 & --- x ---\nl
NGC 5135 & 13 25 44.0 & $-$29 50 00.2 & 12.88 & 0.0132 & 52.9 & 0.256 &
Sy 2 & 2.6 x 1.8\nl
IRAS 18325-5926 & 18 36 58.3 & $-$59 24 08.2 & 13.20 & 0.0192 & 76.8 & 0.372 &
Sy 2 & --- x ---\nl
IC 4995 & 20 19 59.0 & $-$52 37 18.6 & 14.28 & 0.0163 & 65.2 & 0.316 &
Sy 2 & 1.2 x 0.8\nl
IC 5063 & 20 52 02.1 & $-$57 04 06.7 & 12.89 & 0.0113 & 45.2 & 0.219 &
Sy 2 & 2.1 x 1.4\nl
NGC 7130 & 21 48 19.5 & $-$34 57 04.0 & 12.98 & 0.0159 & 63.6 & 0.308 &
Sy 2 & 1.5 x 1.4\nl
NGC 7213 & 22 09 16.2 & $-$47 10 00.2 & 11.01 & 0.0050 & 20.0 & 0.097 &
Sy 1 & 3.1 x 2.8\nl
\enddata

\tablenotetext{a}{3 cm radio positions; except NGC 5135 which is 13 cm}
\tablenotetext{b}{Taken from the NED database}
\tablenotetext{c}{taken from de Grijp et al. (1992), except NGC 7130 which
is taken from van den Broek et al. (1991) and NGC 5135 which is taken from
Phillips et al. (1983)}
\tablenotetext{d}{Based on a Hubble constant of $H_{\rm o}$ = 75 km s$^{-1}$
Mpc$^{-1}$}
\tablenotetext{e}{Typically Seyferts are typed as follows: emission line
galaxies which contain significantly broader (by $\sim$500 km s$^{-1}$)
permitted lines than the forbidden narrow lines are classified as
Seyfert 1 (Sy1), narrow line objects that have 
[OIII]$\lambda$5007/H$\beta$ $>$ 3 and [NII]$\lambda$6584/H$\alpha$
$>$ 0.5 are classified as Seyfert 2 (Sy 2)}
\end{deluxetable}

\clearpage
\begin{deluxetable}{lccc}
\scriptsize
\tablecaption{Clean beam parameters and rms noise}
\tablewidth{0pc}
\tablehead{
\colhead{Name} & \colhead{wavelength} & \colhead{Beam} & \colhead{rms noise}
\\
\colhead{ } & \colhead{(cm)} & \colhead{(arcsec x arcsec)} & 
\colhead{(mJy beam$^{-1}$)}
} 
\startdata
NGC 3393        & 13 & 7.22 x 3.95 & 0.09\nl
                &  6 & 4.07 x 1.91 & 0.10\nl
                &  3 & 2.46 x 1.04 & 0.19\nl
IRAS 11215-2806 & 13 & 8.00 x 3.42 & 0.14\nl
                &  6 & 3.85 x 1.78 & 0.11\nl
                &  3 & 2.12 x 0.99 & 0.19\nl
IRAS 11249-2859 & 13 & 7.85 x 3.17 & 0.04\nl
                &  6 & 3.58 x 1.86 & 0.11\nl 
                &  3 & 2.00 x 1.01 & 0.19\nl
NGC 4507        & 13 & 5.24 x 3.93 & 0.17\nl
                &  6 & 2.58 x 1.98 & 0.10\nl
                &  3 & 1.76 x 1.01 & 0.16\nl
IC 3639         & 13 & 5.33 x 3.86 & 0.05\nl
                &  6 & 2.78 x 2.02 & 0.07\nl
                &  3 & 1.55 x 1.15 & 0.13\nl
IRAS 13059-2407 & 13 & 7.95 x 3.74 & 0.22\nl
                &  6 & 4.35 x 1.93 & 0.27\nl
                &  3 & 2.38 x 1.08 & 0.20\nl
NGC 5135        & 13 & 6.83 x 3.72 & 0.09\nl
IRAS 18325-5926 & 13 & 4.75 x 3.28 & 0.07\nl
                &  6 & 2.35 x 1.55 & 0.08\nl
                &  3 & 1.30 x 0.87 & 0.10\nl
IC 4995         & 13 & 5.01 x 3.16 & 0.07\nl
                &  6 & 2.44 x 1.80 & 0.13\nl
                &  3 & 1.34 x 0.98 & 0.18\nl
IC 5063         & 13 & 4.23 x 3.57 & 0.72\nl
                &  6 & 2.32 x 1.74 & 0.51\nl
                &  3 & 1.31 x 0.97 & 0.16\nl
NGC 7130        & 13 & 6.19 x 3.51 & 0.15\nl
                &  6 & 3.17 x 1.86 & 0.10\nl
                &  3 & 1.71 x 1.01 & 0.11\nl
NGC 7213        & 13 & 4.93 x 3.78 & 0.12\nl
                &  6 & 2.27 x 1.94 & 0.15\nl
                &  3 & 1.26 x 1.08 & 0.16\nl
\enddata
\end{deluxetable}

\clearpage
\begin{deluxetable}{lcccccccc}
\scriptsize
\tablecaption{Radio Properties}
\tablewidth{0pc}
\tablehead{
\colhead{Name} & \colhead{S$_{13cm}$\tablenotemark{a}} & 
\colhead{S$_{6cm}$\tablenotemark{a}} &
\colhead{S$_{3cm}$\tablenotemark{a}} &
\colhead{Log($L_{13cm}$)}&
\colhead{Morphology\tablenotemark{b}} &
\colhead{$\alpha$(6,13)\tablenotemark{c}} &
\colhead{$\alpha$(3,13)\tablenotemark{d}} &
\colhead{$\alpha$(3,6)\tablenotemark{e}}
\\
\colhead { } & \colhead{(mJy)} & \colhead{(mJy)} &
\colhead{(mJy)} & \colhead{W~Hz$^{-1}$} & \colhead{ } & \colhead{ } &
\colhead{ }
} 
\startdata
NGC 3393        & 40.9$\pm$1.1 & 20.5$\pm$0.4 & 13.3$\pm$1.0 & 22.14 &L&  
                  $-$0.91$\pm$0.03 & $-$0.79$\pm$0.02 & $-$0.62$\pm$0.09\nl  
IRAS 11215-2806 & 33.0$\pm$0.7 & 20.0$\pm$0.4 & 12.0$\pm$0.8 & 22.04 &C&
                  $-$0.72$\pm$0.03 & $-$0.67$\pm$0.02 & $-$0.75$\pm$0.07\nl 
IRAS 11249-2859 & 33.4$\pm$0.7 & 22.3$\pm$0.5 & 8.8$\pm$1.0 & 22.52 &L&
                  $-$0.62$\pm$0.04 & $-$0.73$\pm$0.03 & $-$0.79$\pm$0.12\nl 
NGC 4507        & 23.4$\pm$2.0 & 10.8$\pm$0.4 & 8.8$\pm$0.6 & 21.87 &L&
                  $-$0.84$\pm$0.09 & $-$0.59$\pm$0.05 & $-$0.34$\pm$0.11\nl 
IC 3639         & 42.0$\pm$0.5 & 18.2$\pm$0.3 & 9.5$\pm$0.6 & 22.07 &D&
                  $-$0.88$\pm$0.03 & $-$0.92$\pm$0.01 & $-$1.01$\pm$0.06\nl 
IRAS 13059-2407 & 346$\pm$2 & 215$\pm$1 & 108$\pm$1 & 23.10 &C&
                  $-$0.71$\pm$0.01 & $-$0.93$\pm$0.01 & $-$1.17$\pm$0.02\nl
NGC 5135        & 112$\pm$2 & --- & --- & 22.55 &D& --- & --- & ---\nl
IRAS 18325-5926 & 123$\pm$1 & 67.2$\pm$0.4 & 35.5$\pm$0.3 & 22.91 &C&
                  $-$0.88$\pm$0.01 & $-$0.97$\pm$0.01 & $-$1.05$\pm$0.02\nl 
IC 4995         & 8.2$\pm$0.5 & 5.2$\pm$0.5 & 3.1$\pm$0.7 & 21.70 &C&
                  $-$0.61$\pm$0.08 & $-$0.69$\pm$0.05 & $-$0.71$\pm$0.16\nl
IC 5063         & 747$\pm$12 & 389$\pm$5 & 214$\pm$10 & 23.24 &L+D&
                  $-$0.96$\pm$0.02 & $-$0.98$\pm$0.01 & $-$1.05$\pm$0.05\nl 
NGC 7130        & 97.7$\pm$0.7 & 59.0$\pm$0.8 & 30.8$\pm$0.4 & 22.65 &L&
                  $-$0.67$\pm$0.01 & $-$0.84$\pm$0.01 & $-$0.98$\pm$0.01\nl 
NGC 7213        & 168$\pm$1 & 207$\pm$1 & 176$\pm$1 & 21.88 &C&
                  $+$0.30$\pm$0.01 & $+$0.02$\pm$0.01 & $-$0.26$\pm$0.02\nl
\enddata

\tablenotetext{a}{Extended + core flux from BLANKed images}
\tablenotetext{b}{L = Linear Class, D = Diffuse Class, C = Compact Class}
\tablenotetext{c}{Core spectral index for 13 and 6 cm}
\tablenotetext{d}{Core spectral index for 13 and 3 cm}
\tablenotetext{e}{Core spectral index for 6 and 3 cm}
\end{deluxetable}

\clearpage
\begin{deluxetable}{lcccc}
\scriptsize
\tablecaption{Observations made on the sample}
\tablewidth{0pc}
\tablehead{
\colhead{Name} & \colhead{Imaged in radio?}   &
\colhead{Optical long-slit}   &
\colhead{NIR long-slit} & 
\colhead{NIR imaging?}
\\
\colhead{} & \colhead{ } & \colhead{spectra?} & \colhead{spectra?} &
\colhead{ }
} 
\startdata
NGC 3393        & yes & yes & no  & no\nl
IRAS 11215-2806 & yes & no  & no  & no\nl
IRAS 11249-2859 & yes & yes & no  & no\nl
NGC 4507        & yes & no  & no  & no\nl
IC 3639         & yes & yes & no  & no\nl
IRAS 13059-2407 & yes & no  & no  & no\nl
NGC 5135        & yes & yes & no  & no\nl
IRAS 18325-5926 & yes & no  & yes & yes\nl
IC 4995         & yes & no  & yes & yes\nl
IC 5063         & yes & no  & yes & yes\nl
NGC 7130        & yes & no  & no  & yes\nl
NGC 7213        & yes & no  & yes & yes\nl
\enddata
\end{deluxetable}

\clearpage
\begin{deluxetable}{lcccccccccc}
\scriptsize
\tablecaption{Published IRAS Fluxes, and computed q-values}
\tablewidth{0pc}
\tablehead{
\colhead{Name} & \colhead{12$\mu$m}   & \colhead{25$\mu$m}  &
\colhead{60$\mu$m} &  \colhead{100$\mu$m} & \colhead{$\alpha$} &
\colhead{$\alpha$} & \colhead{$\alpha$} &
\colhead{log} & \colhead{q} & \colhead{AGN/starburst}
\\
\colhead{ } & \colhead{(Jy)} & \colhead{(Jy)} & \colhead{(Jy)} &
\colhead{(Jy)} & \colhead{(12,25)} & \colhead{(25,60)} &
\colhead{(60,100)} & \colhead{[FIR/L$\odot$]} &
\colhead{ } & \colhead{composite}
} 
\startdata
NGC 3393&$<$0.25&0.71&2.38&3.94&$<-$1.42&$-$1.38&$-$0.99&10.07&1.92& \nl 
IRAS 11215-2806&$<$0.44&0.31&0.59&$<$1.00&$<$0.48&$-$0.74&$>-$1.03&
$<$9.45&$<$1.41& \nl
IRAS 11249-2859&0.11&0.34&0.60&0.94&$-$1.54&$-$0.65&$-$0.88&9.92&1.40& \nl
NGC 4507&0.46&1.42&4.52&5.40&$-$1.54&$-$1.32&$-$0.35&10.16&2.39&X\nl
IC 3639&0.66&2.32&7.08&10.77&$-$1.71&$-$1.27&$-$0.82&10.33&2.37&X\nl
IRAS 13059-2407&$<$0.25&0.71&1.41&1.75&$<-$1.42&$-$0.78&$-$0.41&9.82&0.72& \nl
NGC 5135&0.64&2.40&16.91&28.64&$-$1.81&$-$2.23&$-$1.03&10.92&2.34&X\nl
IRAS 18325-5926&0.60&1.39&3.17&4.09&$-$1.14&$-$0.94&$-$0.50&10.49&1.53& \nl
IC 4995&$<$0.25&0.36&0.90&1.28&$<-$0.50&$-$1.05&$-$0.69&9.77&2.17&X?\tablenotemark{a}\nl
IC 5063&1.17&3.87&5.89&4.26&$-$1.63&$-$0.48&0.63&10.19&0.94& \nl
NGC 7130&0.59&2.12&16.48&25.57&$-$1.75&$-$2.34&$-$0.86&11.04&2.37&X\nl 
NGC 7213&0.63&0.74&2.56&8.63&$-$0.22&$-$1.42&$-$2.38&9.52&1.43&X\nl
\enddata
\tablenotetext{a}{Inferred from high q-value--see text}
\end{deluxetable}

\clearpage
\begin{deluxetable}{lccccccccc}
\scriptsize
\tablenum{6a}
\tablecaption{IC 3639 reddening-corrected emission-line ratios, relative
to H$\beta$ = 100, at P.A. 90$\arcdeg$ centered on nucleus}
\tablewidth{0pc}
\tablehead{
\colhead{distance\tablenotemark{a}} & 
\colhead{H$\beta$\tablenotemark{b}} & 
\colhead{A$_{v}$\tablenotemark{c}} &
\colhead{~~~[OIII]~~~} &  \colhead{~~~[OIII]~~~} & \colhead{~~~[OI]~~~}  &
\colhead{~~~[NII]~~~}  & \colhead{~~~[NII]~~~} &
\colhead{~~~[SII]~~~}  & \colhead{[SII]}
\\
\colhead{(arcsec)} & \colhead{(x~10$^{-15}$ergs s$^{-1}$ cm$^{-2}$)} &
\colhead{(mag)} &
\colhead{$\lambda$4959} & \colhead{$\lambda$5007} &
\colhead{$\lambda$6300} & \colhead{$\lambda$6548} & 
\colhead{$\lambda$6583} & \colhead{$\lambda$6717} & 
\colhead{$\lambda$6731}
}

\startdata
$-$19.6 &  1.3 & 1.64 & 63.3 & 245 & 30.2 & 23.0 & 105 & 80.9 & 54.0\nl
$-$18.2 &  5.0 & 2.74 &  161 & 361 & 62.4 & 27.1 & 117 & 64.7 & 55.2\nl
$-$16.8 & 61.0 & 1.31 &  241 & 693 & 50.8 & 82.1 & 239 & 57.2 & 63.4\nl
$-$15.4 &  5.3 & 2.04 &  110 & 286 & 29.0 & 36.0 & 110 & 66.1 & 49.0\nl
$-$14.0 &  5.5 & 2.05 &  107 & 290 & 23.3 & 39.7 & 121 & 75.6 & 50.3\nl
$-$12.6 &  3.6 & 1.92 & 99.4 & 305 & 26.4 & 43.8 & 140 & 78.9 & 50.4\nl
$-$11.2 &  2.3 & 1.69 &  127 & 346 &      & 48.9 & 163 & 76.6 & 63.4\nl
$-$9.8  &  6.5 & 2.71 &  194 & 502 & 34.0 & 50.4 & 159 & 82.5 & 49.3\nl
$-$8.4  &  2.5 & 1.55 &  115 & 302 & 34.1 & 48.0 & 156 & 74.2 & 48.4\nl
$-$7.0  &  4.1 & 1.36 & 95.3 & 275 & 23.5 & 47.4 & 145 & 60.0 & 50.4\nl
$-$5.6  &  6.8 & 1.23 & 96.8 & 485 & 29.1 & 64.0 & 166 & 67.4 & 57.9\nl
$-$4.2  & 13.1 & 0.87 &  235 & 656 & 44.0 & 69.5 & 219 & 56.6 & 54.9\nl
$-$2.8  & 61.0 & 1.31 &  241 & 693 & 50.8 & 82.1 & 239 & 57.2 & 63.4\nl
$-$1.4  & 372  & 2.20 &  259 & 745 & 50.5 & 89.0 & 240 & 54.0 & 59.4\nl
   0.0  & 863  & 1.35 &  336 & 891 & 55.4 & 97.3 & 238 & 56.8 & 69.7\nl
$+$1.4  & 54.1 & 3.29 &  201 & 399 & 47.7 & 88.0 & 210 & 64.3 & 52.1\nl
$+$2.8  & 16.9 & 2.45 & 68.6 & 153 & 17.9 & 50.2 & 138 & 57.6 & 42.3\nl
$+$4.2  & 18.1 & 2.49 & 60.2 & 134 & 23.8 & 46.3 & 128 & 59.1 & 42.0\nl
$+$5.6  &  7.3 & 2.27 & 83.6 & 132 & 19.6 & 43.0 & 141 & 67.8 & 51.0\nl
$+$7.0  &  1.9 & 1.24 & 68.6 & 124 & 20.9 & 49.3 & 134 & 80.9 & 57.2\nl
$+$8.4  &  2.1 & 0.68 & 38.9 & 90.4& 12.5 & 36.3 & 109 & 55.8 & 41.0\nl
$+$9.8  &  4.8 & 1.01 & 35.4 & 90.0&  6.2 & 33.9 & 96.2& 47.1 & 36.2\nl
$+$11.2 & 10.6 & 1.85 & 36.1 & 104 &      & 31.8 & 90.5& 45.5 & 35.3\nl
$+$12.6 &  6.0 & 1.85 & 44.8 & 105 & 15.4 & 31.4 & 98.3& 53.1 & 37.0\nl
$+$14.0 &  4.8 & 1.91 & 40.3 & 111 &      & 29.8 & 93.7& 57.1 & 40.1\nl
$+$15.4 &  3.4 & 1.78 & 63.2 & 145 & 21.9 & 28.4 & 98.2& 61.2 & 44.1\nl
$+$16.8 &  4.3 & 2.22 & 64.2 & 233 &      & 27.0 & 97.9& 59.5 & 47.2\nl
$+$18.2 &  2.2 & 1.94 &  101 & 199 & 22.5 & 40.9 & 107 & 70.1 & 46.4\nl
\enddata

\tablenotetext{a}{Negative angular distances are to the east of the
nucleus, positive angular distances are west of the nucleus, and 0.0
indicates the nucleus}
\tablenotetext{b}{H$\alpha$ fluxes may be obtained by multiplying these
fluxes by 2.85}
\tablenotetext{c}{calculated from the Balmer decrement assuming Case B
recombination}
\end{deluxetable}

\clearpage
\begin{deluxetable}{lccccccccc}
\scriptsize
\tablenum{6b}
\tablecaption{IC 3639 reddening-corrected emission-line ratios, relative
to H$\beta$ = 100, at P.A. 90$\arcdeg$, 7$\farcs$5 North of nucleus}
\tablewidth{0pc}
\tablehead{
\colhead{distance\tablenotemark{a}} &
\colhead{H$\beta$\tablenotemark{b}} &
\colhead{A$_{v}$\tablenotemark{c}} & 
\colhead{~~~[OIII]~~~} & \colhead{~~~[OIII]~~~} & \colhead{~~~[OI]~~~} &
\colhead{~~~[NII]~~~} & \colhead{~~~[NII]~~~} &
\colhead{~~~[SII]~~~} & \colhead{~~~[SII]~~~}
\\
\colhead{(arcsec)} & \colhead{(x~10$^{-15}$ergs s$^{-1}$ cm$^{-2}$)} &
\colhead{(mag)} &
\colhead{$\lambda$4959} & \colhead{$\lambda$5007} &
\colhead{$\lambda$6300} & \colhead{$\lambda$6548} & 
\colhead{$\lambda$6583} & \colhead{$\lambda$6717} & 
\colhead{$\lambda$6731}
}

\startdata
$-$15.4 &  4.3 & 1.22 & 49.4 &  136 &  9.7 & 23.8 & 76.2 & 45.7 & 36.4\nl
$-$14.0 &  4.9 & 1.30 & 48.2 &  122 & 16.5 & 26.4 & 81.7 & 48.8 & 36.6\nl
$-$12.6 &  3.3 & 0.94 & 34.0 &  105 & 14.1 & 29.2 & 87.5 & 51.7 & 38.9\nl
$-$11.2 &  3.1 & 0.83 & 33.7 & 91.0 & 14.6 & 29.3 & 91.7 & 55.4 & 37.2\nl
$-$9.8  &  3.6 & 0.72 & 28.7 & 84.4 & 19.2 & 30.6 & 91.1 & 51.8 & 34.5\nl
$-$8.4  &  4.5 & 0.53 & 28.5 & 77.1 & 12.9 & 30.5 & 90.1 & 47.5 & 33.6\nl
$-$7.0  &  5.9 & 0.49 & 31.1 & 81.0 &  9.1 & 28.5 & 85.9 & 44.8 & 31.9\nl
$-$5.6  & 10.6 & 0.91 & 31.0 & 82.2 &  6.4 & 27.4 & 83.8 & 41.3 & 29.2\nl
$-$4.2  & 10.0 & 1.12 & 31.4 & 74.1 &  6.8 & 30.9 & 91.5 & 45.4 & 33.9\nl
$-$2.8  &  8.1 & 1.06 & 29.4 & 80.6 &  9.9 & 33.7 & 96.2 & 49.7 & 35.7\nl
$-$1.4  &  7.9 & 1.00 & 26.9 & 83.6 &  8.2 & 32.8 & 96.3 & 47.7 & 35.8\nl
   0.0  & 14.9 & 1.49 & 30.3 & 72.5 &  5.8 & 31.5 &  100 & 48.9 & 35.6\nl
$+$1.4  & 21.1 & 2.07 & 28.8 & 69.7 &  9.3 & 33.2 &  105 & 52.6 & 38.4\nl
$+$2.8  & 13.3 & 2.11 & 37.1 & 75.9 &      & 36.5 &  108 & 56.7 & 42.3\nl
$+$4.2  &  6.1 & 1.61 & 36.9 & 77.5 &      & 33.0 &  113 & 57.5 & 46.2\nl
$+$5.6  &  4.1 & 1.28 & 29.1 & 67.7 &      & 35.7 &  110 & 61.2 & 43.7\nl
$+$7.0  &  6.4 & 1.86 & 36.0 & 85.3 & 10.2 & 31.6 &  107 & 60.0 & 44.2\nl
$+$8.4  &  4.5 & 1.76 & 34.4 & 71.1 &      & 25.6 &  102 & 65.6 & 41.8\nl
$+$9.8  &  4.7 & 2.12 &      &  123 &      & 30.9 &  106 & 65.2 & 48.5\nl
$+$11.2 &  5.4 & 2.35 &      &  124 &      & 30.9 & 89.3 & 66.2 & 46.4\nl
$+$12.6 &  4.5 & 2.28 &      &  158 &      & 32.2 & 94.7 & 64.0 & 56.0\nl
\enddata

\tablenotetext{a}{Negative angular distances are to the east, positive
angular distances are west, and 0.0 indicates the point 7$\farcs$5 directly
north of the nucleus}
\tablenotetext{b}{H$\alpha$ fluxes may be obtained by multiplying these
fluxes by 2.85}
\tablenotetext{c}{calculated from the Balmer decrement assuming Case B
recombination}
\end{deluxetable}

\clearpage
\begin{deluxetable}{lccccccccc}
\scriptsize
\tablenum{6c}
\tablecaption{IC 3639 reddening-corrected emission-line ratios, relative
to H$\beta$ = 100, at P.A. 0$\arcdeg$ centered on nucleus}
\tablewidth{0pc}
\tablehead{
\colhead{distance\tablenotemark{a}} &
\colhead{H$\beta$\tablenotemark{b}} &
\colhead{A$_{v}$\tablenotemark{c}} & 
\colhead{~~~[OIII]~~~} & \colhead{~~~[OIII]~~~} & \colhead{~~~[OI]~~~} &
\colhead{~~~[NII]~~~} & \colhead{~~~[NII]~~~} &
\colhead{~~~[SII]~~~} & \colhead{~~~[SII]~~~}
\\
\colhead{(arcsec)} & \colhead{(x~10$^{-15}$ergs s$^{-1}$ cm$^{-2}$)} &
\colhead{(mag)} &
\colhead{$\lambda$4959} & \colhead{$\lambda$5007} &
\colhead{$\lambda$6300} & \colhead{$\lambda$6548} & 
\colhead{$\lambda$6583} & \colhead{$\lambda$6717} & 
\colhead{$\lambda$6731}
}
\startdata
$-$19.6 &  1.1 & 1.78 & 65.1 &  151 &      & 35.7 & 90.4 & 90.4 & 41.7\nl
$-$18.2 &  1.5 & 1.44 & 59.2 &  192 & 18.3 & 35.6 & 96.6 & 82.0 & 49.3\nl
$-$16.8 &  2.8 & 1.51 & 61.1 &  139 & 11.4 & 27.7 & 94.9 & 56.7 & 45.8\nl
$-$15.4 &  3.8 & 1.63 & 55.0 & 98.4 & 13.3 & 29.4 & 94.7 & 57.7 & 38.6\nl
$-$14.0 &  4.9 & 1.83 & 45.6 &  115 &  9.8 & 27.7 &  100 & 57.0 & 41.5\nl
$-$12.6 &  4.6 & 1.61 & 34.7 & 97.8 & 12.1 & 30.0 &  101 & 55.8 & 42.7\nl
$-$11.2 &  4.6 & 1.47 & 49.9 & 84.3 &      & 33.1 &  102 & 51.6 & 37.8\nl
$-$9.8  &  8.0 & 1.98 &      & 84.8 &  8.7 & 29.2 &  107 & 52.8 & 34.7\nl
$-$8.4  & 17.0 & 2.50 &      &  112 & 20.9 & 37.4 &  122 & 47.4 & 38.1\nl
$-$7.0  & 20.6 & 2.43 & 58.3 &  140 & 21.5 & 40.3 &  131 & 55.8 & 41.4\nl
$-$5.6  & 25.4 & 2.55 &  164 &  422 & 31.4 & 56.3 &  173 & 58.3 & 51.2\nl
$-$4.2  & 59.9 & 2.46 &  302 &  810 & 50.3 & 86.5 &  228 & 60.3 & 58.4\nl
$-$2.8  &  111 & 1.88 &  262 &  734 & 50.8 & 93.7 &  243 & 57.6 & 59.5\nl
$-$1.4  &  317 & 2.02 &  263 &  737 & 51.1 & 92.4 &  241 & 57.7 & 59.6\nl
   0.0  &  189 & 2.72 &  297 &  848 & 58.2 & 89.9 &  237 & 54.9 & 57.3\nl
$+$1.4  & 60.3 & 2.80 &  191 &  608 & 49.1 & 60.4 &  188 & 50.4 & 47.9\nl
$+$2.8  & 22.5 & 2.00 & 60.4 &  139 & 15.1 & 39.0 &  127 & 46.2 & 35.4\nl
$+$4.2  & 17.9 & 1.50 & 24.7 & 74.9 & 13.1 & 34.2 &  105 & 45.4 & 33.8\nl
$+$5.6  & 19.8 & 1.58 & 26.0 & 72.7 &  6.7 & 30.9 & 97.0 & 44.1 & 33.7\nl
$+$7.0  & 20.5 & 1.92 & 30.2 & 91.2 &  9.3 & 30.1 & 92.2 & 44.1 & 34.0\nl
$+$8.4  & 11.4 & 1.84 & 36.1 &  108 & 32.8 & 31.3 & 92.1 & 47.2 & 34.6\nl
$+$9.8  &  8.4 & 1.54 & 39.0 &  101 & 12.2 & 28.5 & 87.4 & 45.4 & 34.7\nl
$+$11.2 &  7.3 & 1.16 & 40.9 &  118 &  8.0 & 26.7 & 82.6 & 43.5 & 33.3\nl
$+$12.6 &  9.6 & 1.23 & 49.9 &  135 &  8.5 & 23.5 & 75.9 & 45.2 & 32.7\nl
$+$14.0 &  8.2 & 1.16 & 44.3 &  131 & 13.2 & 21.4 & 76.5 & 43.8 & 31.6\nl
$+$15.4 &  4.9 & 1.34 & 34.6 &  133 & 18.8 & 23.1 & 77.5 & 51.3 & 34.2\nl
$+$16.8 &  2.2 & 1.68 & 47.3 &  132 &      & 20.6 & 84.6 & 58.9 & 38.9\nl
\enddata

\tablenotetext{a}{Negative angular distances are to the south of the
nucleus, positive angular distances are north of the nucleus, and 0.0
indicates the nucleus}
\tablenotetext{b}{H$\alpha$ fluxes may be obtained by multiplying these
fluxes by 2.85}
\tablenotetext{c}{calculated from the Balmer decrement assuming Case B
recombination}
\end{deluxetable}

\clearpage
\begin{deluxetable}{lccccccccc}
\scriptsize
\tablenum{7}
\tablecaption{NGC 5135 reddening-corrected emission-line ratios, relative
to H$\beta$ = 100, at P.A. 19$\arcdeg$ centered on nucleus}
\tablewidth{0pc}
\tablehead{
\colhead{distance\tablenotemark{a}} &
\colhead{H$\beta$\tablenotemark{b}} & 
\colhead{A$_{v}$\tablenotemark{c}} &
\colhead{~~~[OIII]~~~} & \colhead{~~~[OIII]~~~} & \colhead{~~~[OI]~~~} &
\colhead{~~~[NII]~~~} & \colhead{~~~[NII]~~~} &
\colhead{~~~[SII]~~~} & \colhead{~~~[SII]~~~}
\\
\colhead{(arcsec)} & \colhead{(x~10$^{-15}$ergs s$^{-1}$ cm$^{-2}$)} &
\colhead{(mag)} &
\colhead{$\lambda$4959} & \colhead{$\lambda$5007} &
\colhead{$\lambda$6300} & \colhead{$\lambda$6548} & 
\colhead{$\lambda$6583} & \colhead{$\lambda$6717} & 
\colhead{$\lambda$6731}
}
\startdata
$-$2.8 &  394 & 4.10 &  268 & 647 & 18.3 & 73.4 & 198 & 40.9 & 36.8\nl
$-$1.4 &  203 & 2.68 &  205 & 572 & 16.9 & 86.2 & 222 & 43.9 & 38.3\nl
   0.0 &  140 & 1.96 &  154 & 409 & 22.4 & 97.9 & 253 & 50.9 & 41.7\nl
$+$1.4 &  125 & 1.87 &  146 & 408 & 24.0 &  110 & 281 & 61.5 & 48.2\nl
$+$2.8 & 89.6 & 1.85 &  116 & 324 & 27.5 &  110 & 287 & 66.5 & 49.3\nl
$+$4.2 & 37.6 & 1.76 & 85.9 & 232 & 29.5 & 43.5 & 114 & 27.5 & 21.0\nl
$+$5.6 & 23.6 & 2.19 & 61.0 & 167 & 14.5 & 64.4 & 169 & 47.0 & 40.7\nl
\enddata

\tablenotetext{a}{Negative angular distances are to the south-west of
the nucleus, positive angular distances are north-east of the nucleus, 
and 0.0 indicates the nucleus}
\tablenotetext{b}{H$\alpha$ fluxes may be obtained by multiplying these
fluxes by 2.85}
\tablenotetext{c}{calculated from the Balmer decrement assuming Case B
recombination}
\end{deluxetable}

%
%

\clearpage
\begin{figure}

\bf{Legends for Figures}

\figurenum{1}
\caption{FIR-radio correlation for sample. Note how the
galaxies lie on the ''radio-bright'' side of the correlation. NGC 4507
has only upper limits for its FIR luminosity.}
\end{figure}

\begin{figure}
\figurenum{2}
\caption{(a) 13 cm radio - optical overlays. The optical
images were obtained from the Digitized Sky Survey (J plates). The
numbers at the top of the overlays are thousands of counts, representing
the stretch for the grey-scale. At the lower right is an inset showing
more clearly the radio structure. (b) 6 and 3 cm radio maps. The
galaxies are presented two by two, from left to right are the lower and
higher frequencies, respectively. In each case, at all frequencies, the
contours are at (-2.5, 2.5, 4, 8, 16, 32, 64, 128, 256, 512, 1024) times
the rms noise (see Table 3 for rms values) measured in a source-free
area of the map.}
\end{figure}

\begin{figure}
\figurenum{3}
\caption{A FIR color-color plot of the IRAS S(100 $\mu$m)/S(12 $\mu$m) 
flux ratio versus the S(60 $\mu$m)/S(25 $\mu$m) ratio. Solid squares
represent the C-class sources, the open square with the circle inside
represents the marginally resolved source (NGC 7213), the crosses
represent the L-class sources, and the open circles represent the
D-class sources. In the case of IRAS 11215-2806, only the S(60) and
S(25) fluxes are known, and the unconstrained S(100)/S(12) ratio is
denoted by an open square with a vertical line running through it.
Interestingly, there is a segregation of the morphological classes in the
color-color plot: the compact sources lie in the lower left (warm FIR
colors) and the diffuse sources (and the marginally resolved source) lie
from the middle to the upper right (cool FIR colors). The L-class
sources span the whole range, but the most highly collimated sources
reside in the lower left.}
\end{figure}

\begin{figure}
\figurenum{4}
\caption{Spectra of IC 3639, PA 90$\arcdeg$, going from AGN (a) to LINER
(b) to H II-region-like (c). Note in that in (b) we marginally detect an 
H$\beta$ absorption feature. Confirmation of this feature is pending.
The vertical scale is F$_{\lambda}$ (ergs cm$^{\rm -2}$ s$^{\rm -1}$
${\rm \AA^{\rm -1}}$) and the wavelengths are in the rest frame of the
galaxy.}
\end{figure}

\begin{figure}
\figurenum{5}
\caption{BPT and VO diagrams for IC 3639, PA 90$\arcdeg$. For this figure,
and for the following BPT and VO diagrams: (a) is log([NII]/H$\alpha$)
vs log([OIII]/H$\beta$), (b) is log([SII]/H$\alpha$) vs
log([OIII]/H$\beta$), (c) is log([OI]/H$\alpha$ vs log([OIII]/H$\beta$),
and the points proceed radially away from the nucleus APPROXIMATELY from
upper right to lower left with 1$\farcs$4 increments. Of interest in these
figures is the east-west asymmetry. To the east of the nucleus, where
there is a rapidly declining radio emission, the optical emission is
consistent with a LINER and to the west, the emission is more like that
of H II-regions. This effect is seen in all three diagnostics.}
\end{figure}

\begin{figure}
\figurenum{6}
\caption{BPT and VO diagrams for IC 3639, PA 0$\arcdeg$. Again, the optical
emission proceeds, radially away from the nucleus, from AGN to
H II-region-like. Note that south of the nucleus the emission proceeds to
H II-region-like at a larger relative distance than to the north. The
emission to the south dips into the LINER region (at $\sim$5$\farcs$6), but
only significantly in the [OI] diagnostic.}
\end{figure}

\begin{figure}
\figurenum{7}
\caption{BPT and VO diagrams for IC 3639, PA 90$\arcdeg$, 7$\farcs$5 North of
the nucleus. This emission is consistent with H II-regions. Of interest
is (c), which mimics the east-west asymmetry seen in Figure 5a,c. To the
east, just north of the nucleus, the emission is consistent with a
LINER, and not until $\sim$5$\farcs$6 east does it dip into the H II-region
portion of the graph.}
\end{figure}

\begin{figure}
\figurenum{8}
\caption{Spectra of NGC 5135, along PA 19$\arcdeg$, going from AGN (a) to 
LINER (b) to H II-region-like (c).  The vertical scale is F$_{\lambda}$
(ergs cm$^{\rm -2}$ s$^{\rm -1}$ ${\rm \AA^{\rm -1}}$) and the
wavelengths are in the rest frame of the galaxy.} 
\end{figure}

\begin{figure}
\figurenum{9}
\caption{BPT and VO diagrams for NGC 5135, PA 19$\arcdeg$. Of interest is a
north-east south-west asymmetry. To the north-east (there is a radio knot
at 13 cm along this position angle, see Fig. 2a) the emission goes from
AGN to LINER to H II-region-like, whereas to the south-west the emission
remains consistent with an AGN.}
\end{figure}

\begin{figure}
\figurenum{10}
\caption{CTIO spectra. (a) Nuclear spectrum for NGC 3393,  along PA 
90$\arcdeg$. (b) Nuclear spectrum for IRAS 11249-2859 along PA
90$\arcdeg$.  The vertical scale is F$_{\lambda}$ (ergs cm$^{\rm -2}$
s$^{\rm -1}$ ${\rm \AA^{\rm -1}}$) and the wavelengths are in the rest frame
of the galaxy.} 
\end{figure}

\begin{figure}
\figurenum{11}
\caption{K-band IR spectra for IRAS 18325-5926, IC 5063, NGC 7213, and
NGC 4995. Note that we have detected H$_{2}$ and Br$\gamma$ in all four
cases. Also detected in all cases, except IRAS 1832, is the CO 2-0
band-head absorption feature.}
\end{figure}

\begin{figure}
\figurenum{12}
\caption{K-band continuum IR images of NGC 7130. Figure 12a shows a
contour plot of the intensity of the K-band continuum emission. It
clearly shows the two tightly wrapped spiral arms that emanate from a
somewhat asymmetric bar. In Figure 12b, the high-contrast grey-scale
image, one can clearly see the ring-like structure that connects with
the companion to the north-west.}
\end{figure}

\begin{figure}
\figurenum{13}
\caption{3 cm radio image of IC 5063 overlayed with the centroid
positions of the H$\alpha$ knots (taken from Heisler \& Vader, 1995)
which are represented by crosses. (a) assumes that radio knot B is the 
nucleus and the H$\alpha$ knot and knot B are made to align. (b) assumes
that radio knot A is the nucleus and the H$\alpha$ knot and knot A are
made to align. The radio contours are 0.16 mJy beam$^{-1}$ x
(-2.5,2.5,3,4,8,10,16,20,25,30,32,40,50,64,70,90,128,256,512,800).}
\end{figure}
\end{document}